\documentclass[11pt]{article}

\usepackage[margin=1in]{geometry}
\usepackage[T1]{fontenc}
\usepackage[utf8]{inputenc}
\usepackage{times}
\usepackage{setspace}
\usepackage{booktabs}
\usepackage{tabularx}
\usepackage{array}
\usepackage{multirow}
\usepackage{graphicx}
\usepackage{enumitem}
\usepackage{amsmath}
\usepackage{natbib}
\usepackage{url}
\usepackage[hidelinks]{hyperref}
\usepackage{xcolor}
\usepackage{tikz}
\usepackage{pgfplots}
\usetikzlibrary{arrows.meta, positioning, shapes.geometric, shapes.misc, fit}
\pgfplotsset{compat=1.18}

\definecolor{accent}{HTML}{1F4E79}
\definecolor{accentsoft}{HTML}{EAF1F7}
\definecolor{linegray}{HTML}{B8C4CF}
\definecolor{textgray}{HTML}{586575}
\definecolor{greenish}{HTML}{2D8B61}
\definecolor{golden}{HTML}{D4862E}
\definecolor{violetish}{HTML}{6A56B2}

\newcolumntype{Y}{>{\raggedright\arraybackslash}X}

\title{\textbf{Lishu:}\\
A Real-Source Research Workbench for Elite Business Journal Search, Analysis, and Writing Support}

\author{
Chuang Zhao \quad Hongke Zhao \quad Yichen Li \quad Xiaoquan Zhi \quad Songyue Guo\\
ADM Lab, \url{https://www.adm-cube.online/}\\
\small
\url{zhaochuang@tju.edu.cn} \quad \url{hongke@tju.edu.cn} \quad \url{ycli0204@hust.edu.cn}\\
\url{zhixiaoquan@tju.edu.cn} \quad \url{sguo349@connect.hkust-gz.edu.cn}
}

\date{}

\begin{document}

\maketitle

\begin{abstract}
This paper presents \textit{Lishu}, a deployable web artifact for searching, monitoring, and interpreting literature from elite business and management journals. The system integrates the UTD-24 and Financial Times 50 (FT50) journal pools and combines Crossref, OpenAlex, Unpaywall, and optional CORE enrichment to support a broader research workflow than article retrieval alone. In the current implementation, users can search across curated journal pools, apply multi-journal filters, preview open full-text excerpts when available, generate citations and exports, inspect topic and affiliation structure, produce review drafts, simulate virtual peer review, and assemble grant-oriented research narratives. Unlike static journal directories or general-purpose academic search engines, the artifact is explicitly scoped to high-status management outlets and is designed to support sensemaking tasks that matter to researchers, doctoral students, and lab managers: identifying recent work, surfacing topical concentration, comparing themes, and converting search output into actionable research material. Architecturally, the system emphasizes source transparency, modularity, and low-cost public deployability through a lightweight Node.js service layer, a multi-page client interface, optional large-language-model enhancement for interpretation and writing support, and a free-tier persistence path through Supabase. The paper contributes both a functioning design artifact and an extensible architectural pattern for journal-pool-specific scholarly discovery and writing support, with implications for digital research infrastructure in information systems and business scholarship.
\end{abstract}

\textbf{Keywords:} design science, scholarly search, research analytics, business journals, Crossref, OpenAlex, digital research infrastructure

\section{Introduction}

Researchers in business schools often rely on elite journal lists as practical filters for literature review, outlet targeting, and academic evaluation. Two such lists are especially influential: the \textbf{UTD-24}, commonly used in research productivity assessment, and the \textbf{FT50}, frequently referenced in ranking, accreditation, and prestige signaling contexts. Yet the working process around these lists remains fragmented. Journal pools are typically maintained separately from live literature search, publisher pages provide inconsistent browsing experiences, and monitoring topical developments across multiple elite journals is labor intensive.

This paper introduces \textit{Lishu}, a web-based research artifact designed to reduce that fragmentation. The artifact does not attempt to replace publisher websites or commercial bibliographic databases. Rather, it provides a focused, domain-specific decision support layer that combines journal-pool curation, real-source metadata retrieval, full-text availability enrichment, topic extraction, export utilities, review-generation support, and lightweight behavioral analytics. In design-science terms, the artifact addresses a class of recurrent scholarly search and sensemaking problems by creating a practical digital object for use in a specific problem domain \citep{hevner2004, gregor2013}. The broader ambition is to show that even a relatively lightweight artifact can offer meaningful infrastructure value when its scope, sourcing logic, and task support are carefully aligned.

The current public instance of the artifact is available online\footnote{\url{https://utd24-journal-portal.onrender.com/}}, which allows the system to function not only as a conceptual design exemplar but also as a live, inspectable research artifact.

The artifact was designed around three overarching objectives. First, it should help researchers quickly search within a trusted set of top management and business journals. Second, it should provide interpretable signals about emerging topics from returned results rather than merely displaying lists of papers. Third, it should remain easy to deploy and maintain in low-budget, lab-managed, or publicly hosted environments.

The paper proceeds as follows. Section~\ref{sec:background} positions the artifact in design science and scholarly discovery work. Section~\ref{sec:objectives} outlines design objectives and artifact requirements. Section~\ref{sec:architecture} presents the architecture and search-to-insight workflow. Section~\ref{sec:analytics} describes the research analytics layer. Section~\ref{sec:implementation} discusses implementation and deployment decisions. The final sections discuss contributions, limitations, and future directions.

\section{Background and Research Motivation}
\label{sec:background}

Design science research in information systems emphasizes the creation and evaluation of artifacts that address relevant organizational and human problems \citep{hevner2004}. In this tradition, the artifact presented here is best understood as a domain-specific search and sensemaking system for scholarly work. It occupies the intersection of information retrieval, research support, and digital infrastructure for academic decision making. The design also resonates with the view that information systems artifacts become more valuable when they embed both operational functionality and interpretive support \citep{march1995, gregor2013}.

The motivating problem is not a lack of journal information per se. Instead, the problem lies in the separation between (1) curated notions of journal quality and prestige, (2) real-time metadata about recent papers, and (3) tools that help scholars quickly interpret what a search result set means. General search engines offer breadth, but they are not optimized for journal-pool-specific exploration. Static journal directories offer curation, but rarely offer live query-based insight generation. This artifact addresses that gap through a lightweight but integrated workflow. From a research-infrastructure perspective, the artifact is promising because it reduces the translation cost between ``finding papers'' and ``understanding what a query reveals about a field.''

This positioning is also consistent with prior work that compares scholarly search and citation infrastructures across coverage, retrieval quality, and analytical usefulness. Comparative studies have shown that Google Scholar, Web of Science, Scopus, PubMed, and other academic search systems differ materially in coverage, precision, transparency, and suitability for specific scholarly tasks \citep{bakkalbasi2006, falagas2008, harzing2016, martinmartin2018, gusenbauer2022}. In parallel, scientometric work has demonstrated the value of visual and structural techniques for detecting emerging topics and research fronts in large document collections \citep{chen2006}. The present artifact contributes to this broader conversation by operationalizing a journal-pool-specific layer for management research rather than attempting universal scholarly search.

\begin{table}[t]
\centering
\caption{Problem Context and Artifact Response}
\label{tab:problem-response}
\begin{tabularx}{\textwidth}{>{\raggedright\arraybackslash}p{0.26\textwidth} >{\raggedright\arraybackslash}p{0.33\textwidth} Y}
\toprule
\textbf{Problem Area} & \textbf{Observed Limitation} & \textbf{Artifact Response} \\
\midrule
Elite journal navigation & UTD-24 and FT50 are consulted separately; outlet information is fragmented & Unified journal-pool directory with pool membership, official links, and submission links \\
Current literature discovery & Researcher must scan multiple journal or publisher pages manually & Live query-based retrieval from Crossref constrained by journal pools \\
Topic sensing & Search results alone do not reveal topical concentration or method signals & Hotspot extraction, distributions, and result-set summaries \\
Practical research workflow & Search, export, citation formatting, and bookmarking are disconnected & Citation modal, favorites, DOI links, and CSV/report export \\
Public deployment and maintenance & Low-cost hosting often breaks persistence and increases maintenance burden & Minimal Node.js stack with optional Supabase-backed persistence \\
\bottomrule
\end{tabularx}
\end{table}

\section{Design Objectives}
\label{sec:objectives}

The artifact was built around five design objectives.

\textbf{Domain focus.} The portal is intentionally limited to journals in the UTD-24 and FT50 pools. This restriction reduces noise and aligns the interface with business-school research evaluation logic.

\textbf{Real-source transparency.} All journal and article content is tied to identifiable sources. Journal pool membership is documented, article metadata are retrieved from Crossref, and output records expose DOI and source links.

\textbf{Query-to-insight workflow.} The system should not only retrieve search results but also help users interpret them through hotspot analysis, distributions, affiliation views, and simple trend analytics.

\textbf{Practical research support.} Citation export, favorites, downloadable CSV output, full-text preview, paper interpretation, and structured writing assistance are included because they reduce friction in everyday research tasks.

\textbf{Low-cost deployability.} The artifact should remain easy to deploy without a heavy dependency chain, while still supporting persistent analytics on a free hosting path.

\begin{table}[t]
\centering
\caption{Core Design Objectives and Implemented Features}
\label{tab:design-objectives}
\begin{tabularx}{\textwidth}{>{\raggedright\arraybackslash}p{0.22\textwidth} >{\raggedright\arraybackslash}p{0.33\textwidth} Y}
\toprule
\textbf{Design Objective} & \textbf{Operationalization} & \textbf{Implemented Features} \\
\midrule
Domain focus & Restrict scope to elite business journals & UTD-24/FT50 pools, list filters, journal directory \\
Source transparency & Expose provenance at the object level & DOI links, source notes, official site links, submission links \\
Search-to-insight & Transform query results into interpretable summaries & Hotspots, distributions, affiliation ranking, method signals \\
Workflow support & Reduce post-search friction & Citation modal, BibTeX export, favorites, CSV export \\
Deployability & Keep infrastructure lean and maintainable & Node.js server, static front end, Render deployment, Supabase persistence \\
\bottomrule
\end{tabularx}
\end{table}

\section{Artifact Overview}

The artifact follows a lightweight client-server architecture. The front end is a modular multi-page website; the back end is a small Node.js application that serves static assets, proxies metadata retrieval, computes analytics, and persists usage events. The current implementation is intended as a publicly deployable research artifact rather than a large-scale platform. This design choice is deliberate: the goal is to demonstrate that a comparatively small technical stack can still support a meaningful research workflow when the architecture is organized around clear scholarly tasks and traceable data provenance.

\subsection{Front-End Structure}

The user-facing interface is organized into five modules:

\begin{enumerate}[leftmargin=1.3em]
\item \textbf{Home}: overview, source note, and quick-access entry points.
\item \textbf{Journal Directory}: filtered browsing of UTD-24 and FT50 journals.
\item \textbf{Article Search}: multi-source article retrieval, filters, full-text access, hotspot analysis, export, favorites, and citation support.
\item \textbf{Comparative and Daily Views}: topic comparison, daily top-paper briefing, and author leaderboard analytics.
\item \textbf{Research Writing Tools}: review generation, virtual review, paper reading, grant-proposal drafting, and AutoResearch orchestration.
\item \textbf{Resources and Analytics}: conferences, submission tools, usage analytics, and support materials.
\end{enumerate}

\subsection{Back-End Services}

The Node.js back end performs four roles:

\begin{enumerate}[leftmargin=1.3em]
\item static file serving,
\item metadata retrieval, enrichment, and normalization,
\item analytical summary generation,
\item writing-support workflow orchestration,
\item usage analytics storage and reporting.
\end{enumerate}

\begin{table}[t]
\centering
\caption{Front-End and Back-End Technical Composition}
\label{tab:technical-stack}
\begin{tabularx}{\textwidth}{>{\raggedright\arraybackslash}p{0.22\textwidth} >{\raggedright\arraybackslash}p{0.27\textwidth} Y}
\toprule
\textbf{Layer} & \textbf{Implemented Technology} & \textbf{Technical Role in the Artifact} \\
\midrule
Presentation layer & HTML, CSS, vanilla JavaScript & Multi-page interface for search, journal browsing, writing support, export utilities, and analytics views \\
Search interaction & Browser-side form state and query serialization & Stable query submission, multi-journal filtering, shareable URLs, local favorites, and citation actions \\
Visualization layer & ECharts & Hotspot ranking, distributional views, and cumulative traffic analytics \\
Application layer & Node.js HTTP server & Static asset delivery, API routing, metadata orchestration, writing workflow jobs, and result normalization \\
Metadata sources & Crossref REST API + OpenAlex & Live retrieval of paper-level metadata, dates, authors, affiliations, citations, and DOI records \\
Full-text enrichment & Unpaywall + optional CORE & Open-access location resolution, PDF availability, and excerpt-level preview support \\
Persistence layer & Supabase or local JSON fallback & Analytics durability across deployments and lightweight operational monitoring \\
Deployment layer & Render-compatible hosting path & Public accessibility with minimal infrastructure overhead \\
\bottomrule
\end{tabularx}
\end{table}

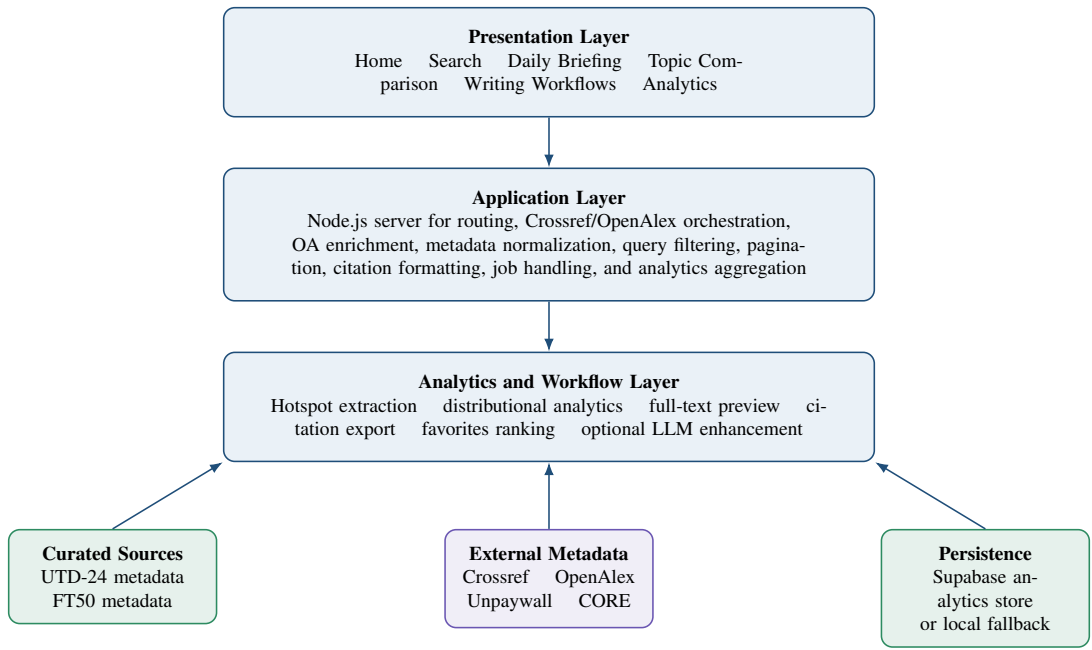
\begin{figure}[t]
\centering
\resizebox{0.88\textwidth}{!}{
\begin{tikzpicture}[
  font=\small,
  node distance=9mm and 10mm,
  layer/.style={draw=accent, rounded corners=2mm, thick, fill=accentsoft, align=center, minimum height=10mm, text width=0.66\textwidth, inner sep=4mm},
  source/.style={draw=greenish, rounded corners=2mm, thick, fill=greenish!12, align=center, minimum height=9mm, text width=0.19\textwidth, inner sep=3mm},
  service/.style={draw=violetish, rounded corners=2mm, thick, fill=violetish!10, align=center, minimum height=9mm, text width=0.19\textwidth, inner sep=3mm},
  arrow/.style={-Latex, thick, draw=accent}
]
\node[layer] (frontend) {\textbf{Presentation Layer}\\Home \quad Search \quad Daily Briefing \quad Topic Comparison \quad Writing Workflows \quad Analytics};
\node[layer, below=of frontend] (backend) {\textbf{Application Layer}\\Node.js server for routing, Crossref/OpenAlex orchestration, OA enrichment, metadata normalization, query filtering, pagination, citation formatting, job handling, and analytics aggregation};
\node[layer, below=of backend] (analysis) {\textbf{Analytics and Workflow Layer}\\Hotspot extraction \quad distributional analytics \quad full-text preview \quad citation export \quad favorites ranking \quad optional LLM enhancement};
\node[source, below left=12mm and 1mm of analysis] (seed) {\textbf{Curated Sources}\\UTD-24 metadata\\FT50 metadata};
\node[service, below=12mm of analysis] (crossref) {\textbf{External Metadata}\\Crossref \quad OpenAlex\\Unpaywall \quad CORE};
\node[source, below right=12mm and 1mm of analysis] (supabase) {\textbf{Persistence}\\Supabase analytics store\\or local fallback};

\draw[arrow] (frontend) -- (backend);
\draw[arrow] (backend) -- (analysis);
\draw[arrow] (seed.north) -- (analysis.south west);
\draw[arrow] (crossref.north) -- (analysis.south);
\draw[arrow] (supabase.north) -- (analysis.south east);
\end{tikzpicture}
}
\caption{System architecture of the Lishu research workbench artifact}
\label{fig:architecture}
\end{figure}

\section{Search-to-Insight Workflow}
\label{sec:architecture}

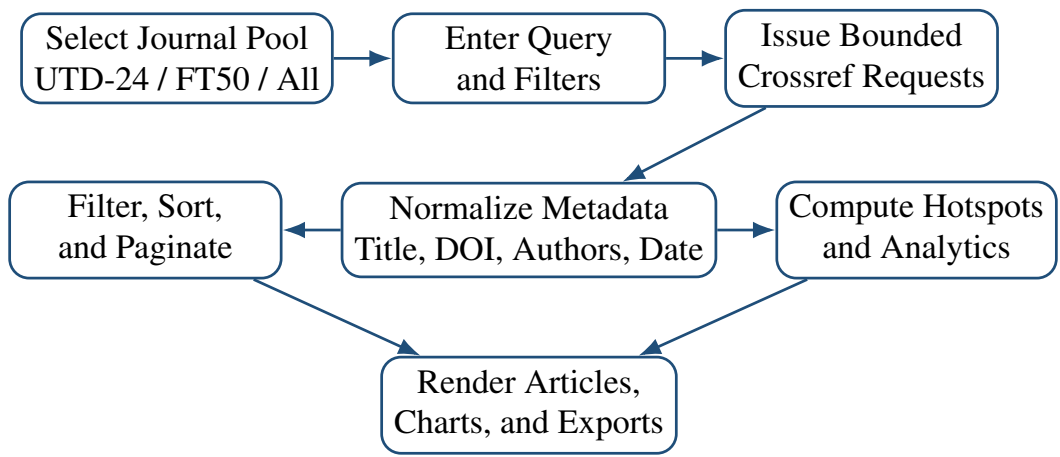
\begin{figure}[t]
\centering
\resizebox{0.86\textwidth}{!}{
\begin{tikzpicture}[
  font=\small,
  node distance=8mm and 6mm,
  flowstep/.style={draw=accent, rounded corners=2mm, thick, fill=white, align=center, minimum height=9mm, minimum width=28mm},
  arrow/.style={-Latex, thick, draw=accent}
]
\node[flowstep] (q1) {Select Journal Pool\\UTD-24 / FT50 / All};
\node[flowstep, right=of q1] (q2) {Enter Query\\and Filters};
\node[flowstep, right=of q2] (q3) {Issue Bounded\\Crossref Requests};
\node[flowstep, below=of q2] (q4) {Normalize Metadata\\Title, DOI, Authors, Date};
\node[flowstep, left=of q4] (q5) {Filter, Sort,\\and Paginate};
\node[flowstep, right=of q4] (q6) {Compute Hotspots\\and Analytics};
\node[flowstep, below=of q4] (q7) {Render Articles,\\Charts, and Exports};

\draw[arrow] (q1) -- (q2);
\draw[arrow] (q2) -- (q3);
\draw[arrow] (q3) -- (q4);
\draw[arrow] (q4) -- (q5);
\draw[arrow] (q4) -- (q6);
\draw[arrow] (q5) -- (q7);
\draw[arrow] (q6) -- (q7);
\end{tikzpicture}
}
\caption{Search-to-insight workflow}
\label{fig:workflow}
\end{figure}

The portal implements a query-to-insight workflow rather than a search-only pipeline. Figure~\ref{fig:workflow} summarizes the operational sequence.

Given a keyword and optional filters, the back end first identifies the journals that belong to the selected pool. It then issues bounded Crossref requests journal by journal, supplements results with OpenAlex records when available, enriches access metadata through Unpaywall and optional CORE retrieval, normalizes the returned data into a common article schema, applies filtering and sorting logic, and computes analytical summaries from the resulting set.

This workflow is important because it turns journal-pool curation into an operational search boundary. Rather than querying a global metadata space and later attempting to infer whether results belong to top journals, the system begins from a trusted elite-journal pool and then builds the result set upward.

\begin{figure}[t]
\centering
\includegraphics[width=0.82\textwidth,height=0.7\textheight,keepaspectratio]{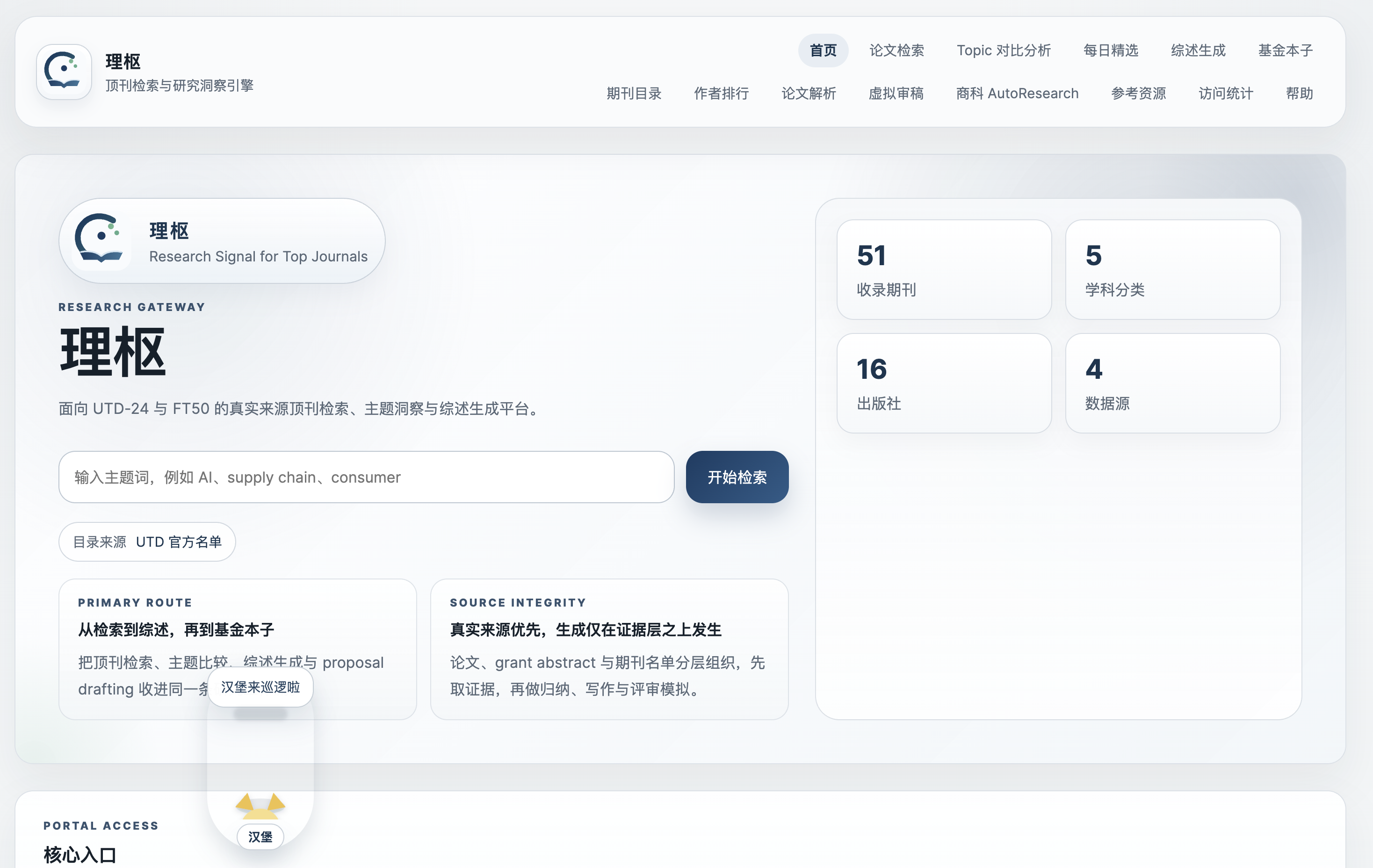}
\caption{Implemented homepage view showing real-source positioning, quick access modules, and statistics}
\label{fig:homepage-screenshot}
\end{figure}

\begin{figure}[t]
\centering
\includegraphics[width=0.82\textwidth,height=0.7\textheight,keepaspectratio]{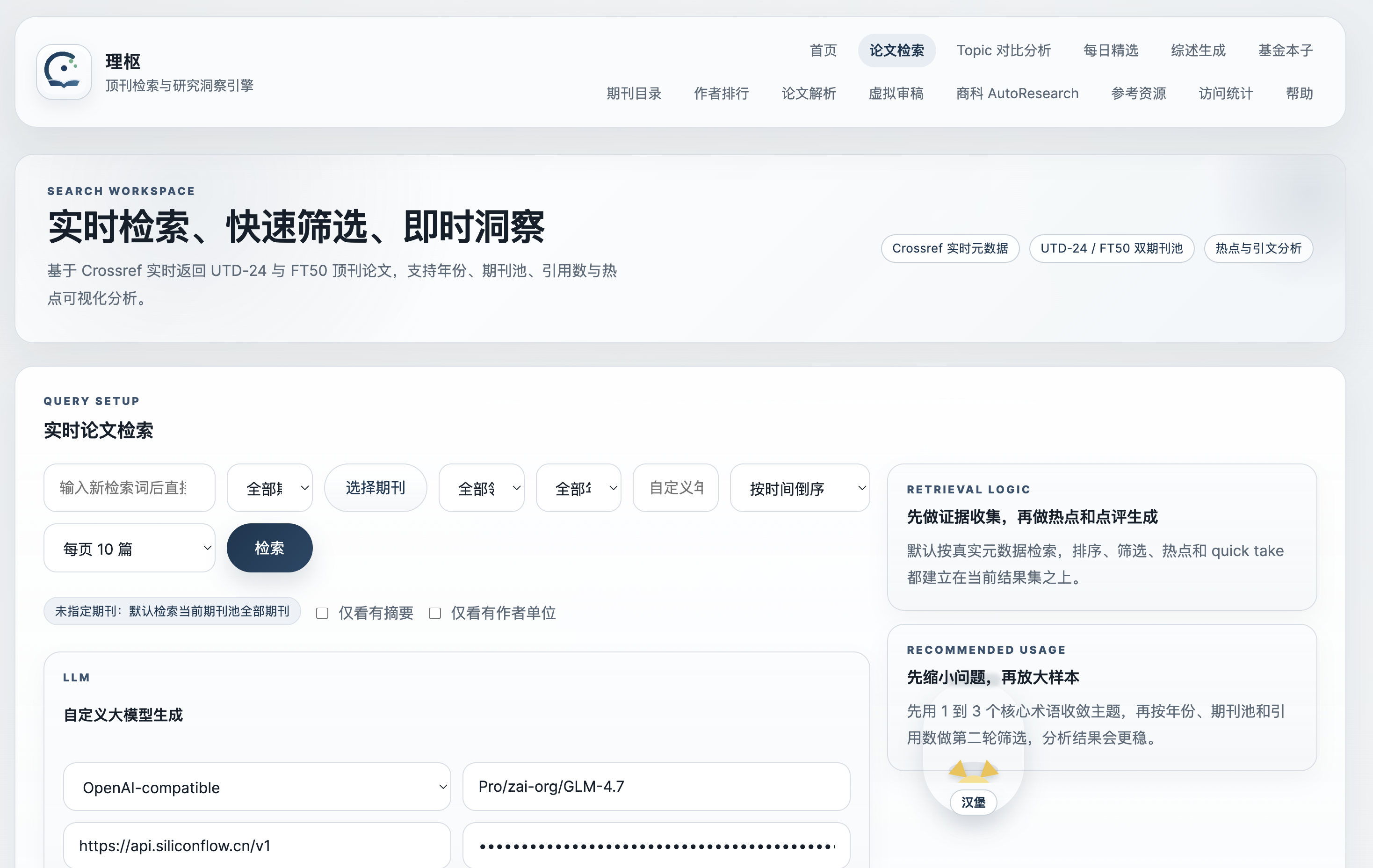}
\caption{Implemented retrieval workspace showing live article search, multi-journal filtering, full-text preview entry points, and research analytics}
\label{fig:search-screenshot}
\end{figure}

\begin{figure}[t]
\centering
\begin{minipage}[t]{0.49\textwidth}
  \centering
  \includegraphics[width=\textwidth,height=0.34\textheight,keepaspectratio]{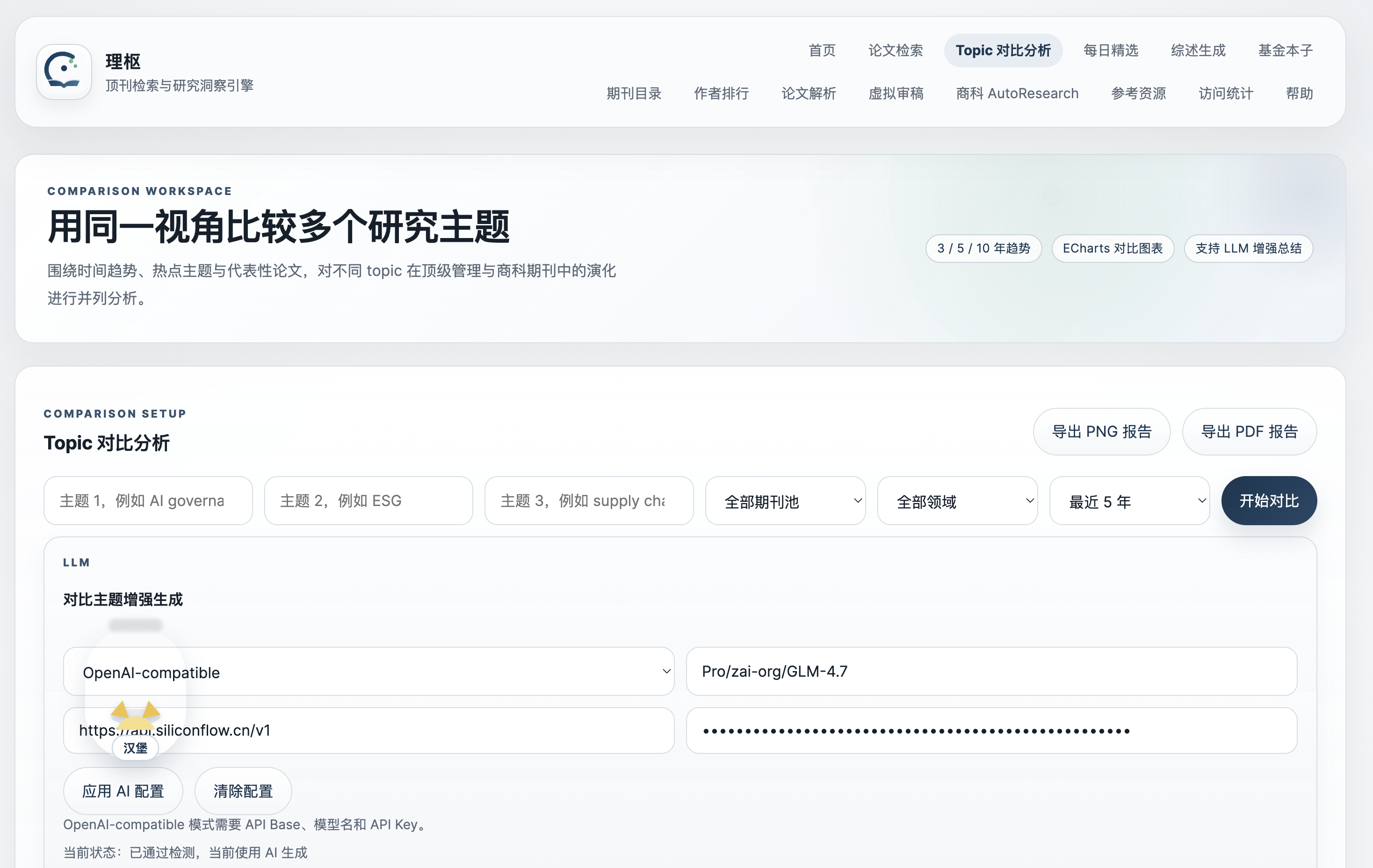}
\end{minipage}
\hfill
\begin{minipage}[t]{0.49\textwidth}
  \centering
  \includegraphics[width=\textwidth,height=0.34\textheight,keepaspectratio]{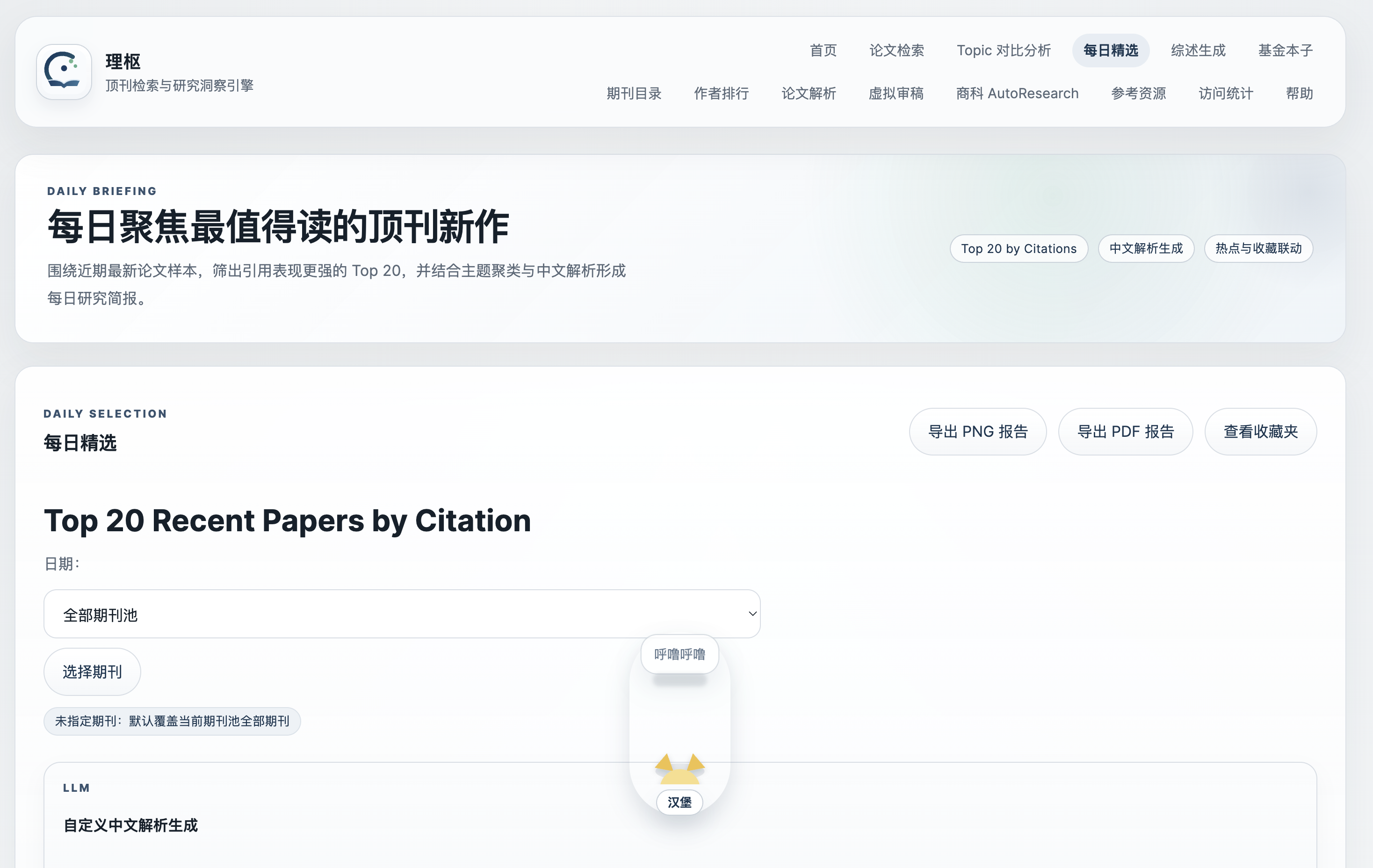}
\end{minipage}
\caption{Implemented comparison and daily-briefing views showing topic timeline analysis and top-paper briefing workflows}
\label{fig:comparison-daily-screenshot}
\end{figure}

\begin{figure}[t]
\centering
\begin{minipage}[t]{0.49\textwidth}
  \centering
  \includegraphics[width=\textwidth,height=0.34\textheight,keepaspectratio]{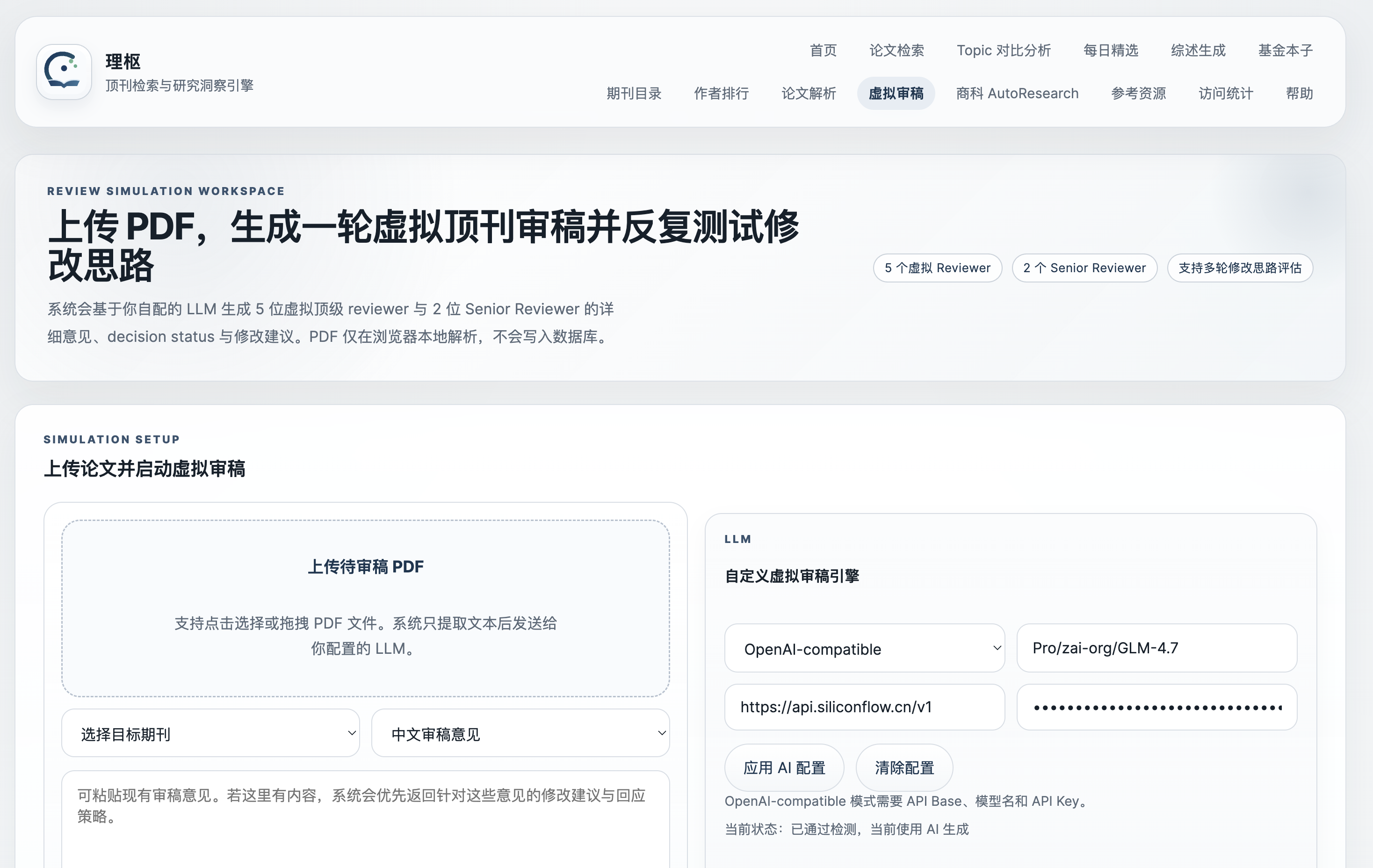}
\end{minipage}
\hfill
\begin{minipage}[t]{0.49\textwidth}
  \centering
  \includegraphics[width=\textwidth,height=0.34\textheight,keepaspectratio]{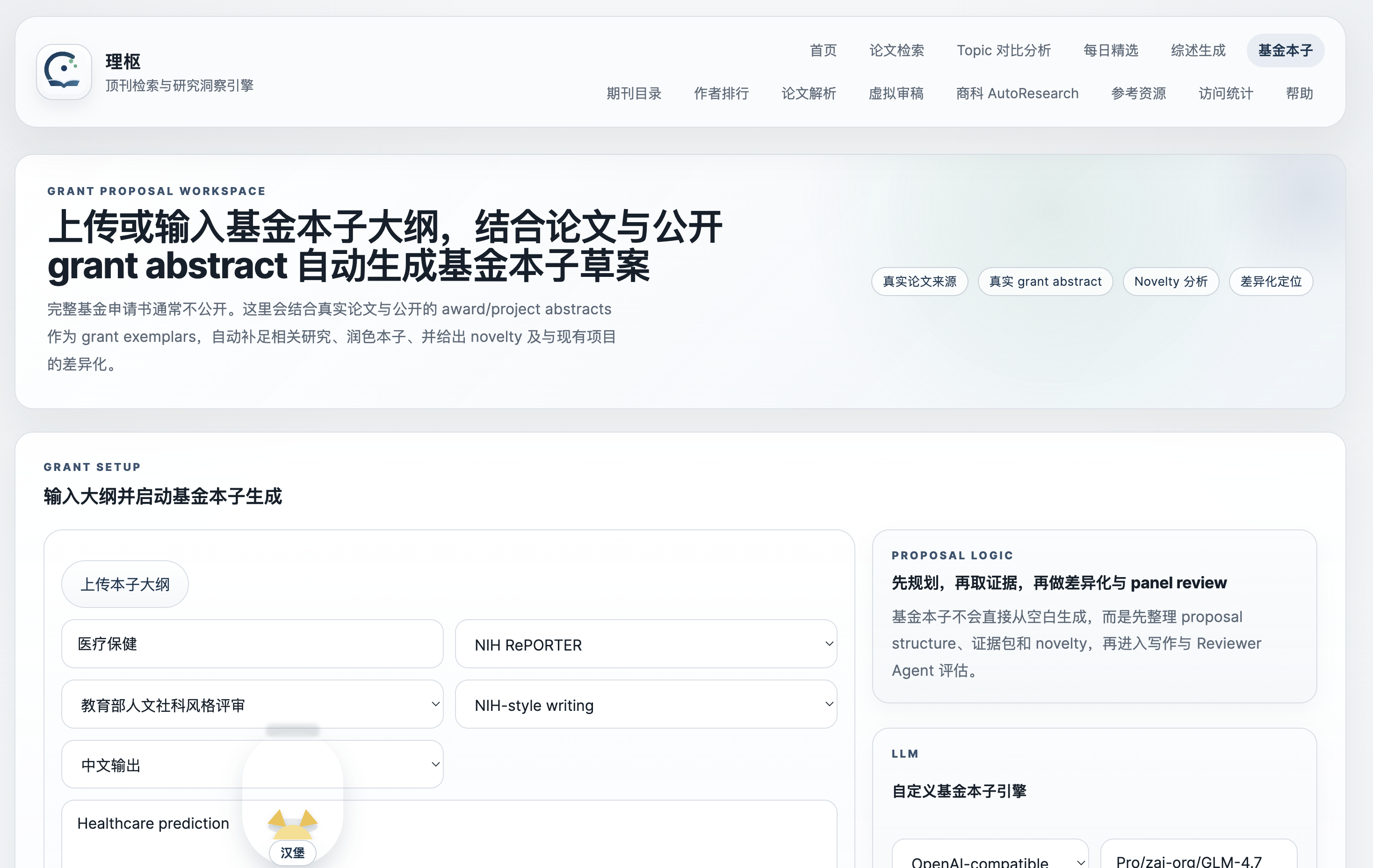}
\end{minipage}
\caption{Implemented writing-support views showing review generation and grant-proposal composition workflows}
\label{fig:writing-screenshot}
\end{figure}

\begin{figure}[t]
\centering
\begin{minipage}[t]{0.49\textwidth}
  \centering
  \includegraphics[width=\textwidth,height=0.34\textheight,keepaspectratio]{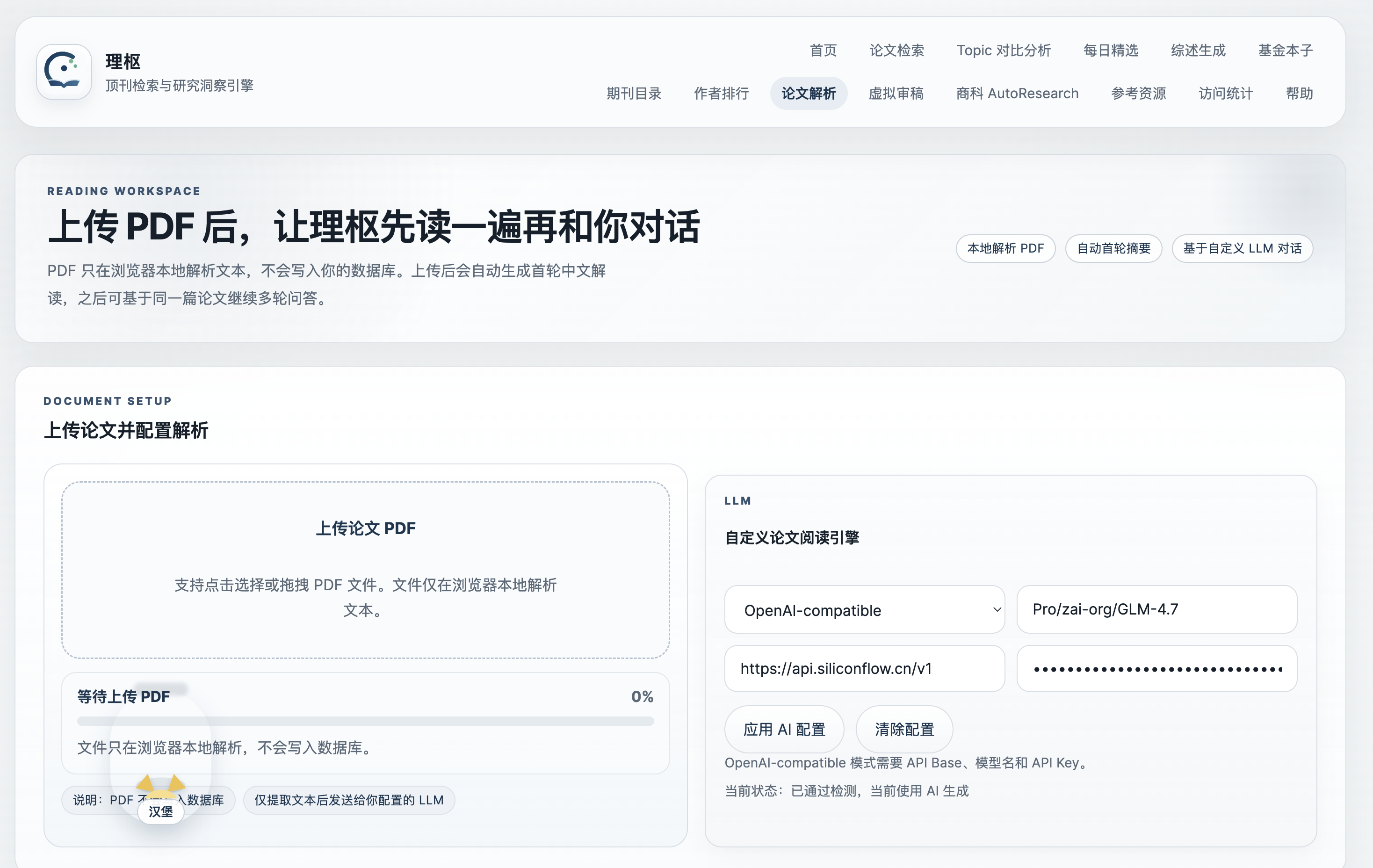}
\end{minipage}
\hfill
\begin{minipage}[t]{0.49\textwidth}
  \centering
  \includegraphics[width=\textwidth,height=0.34\textheight,keepaspectratio]{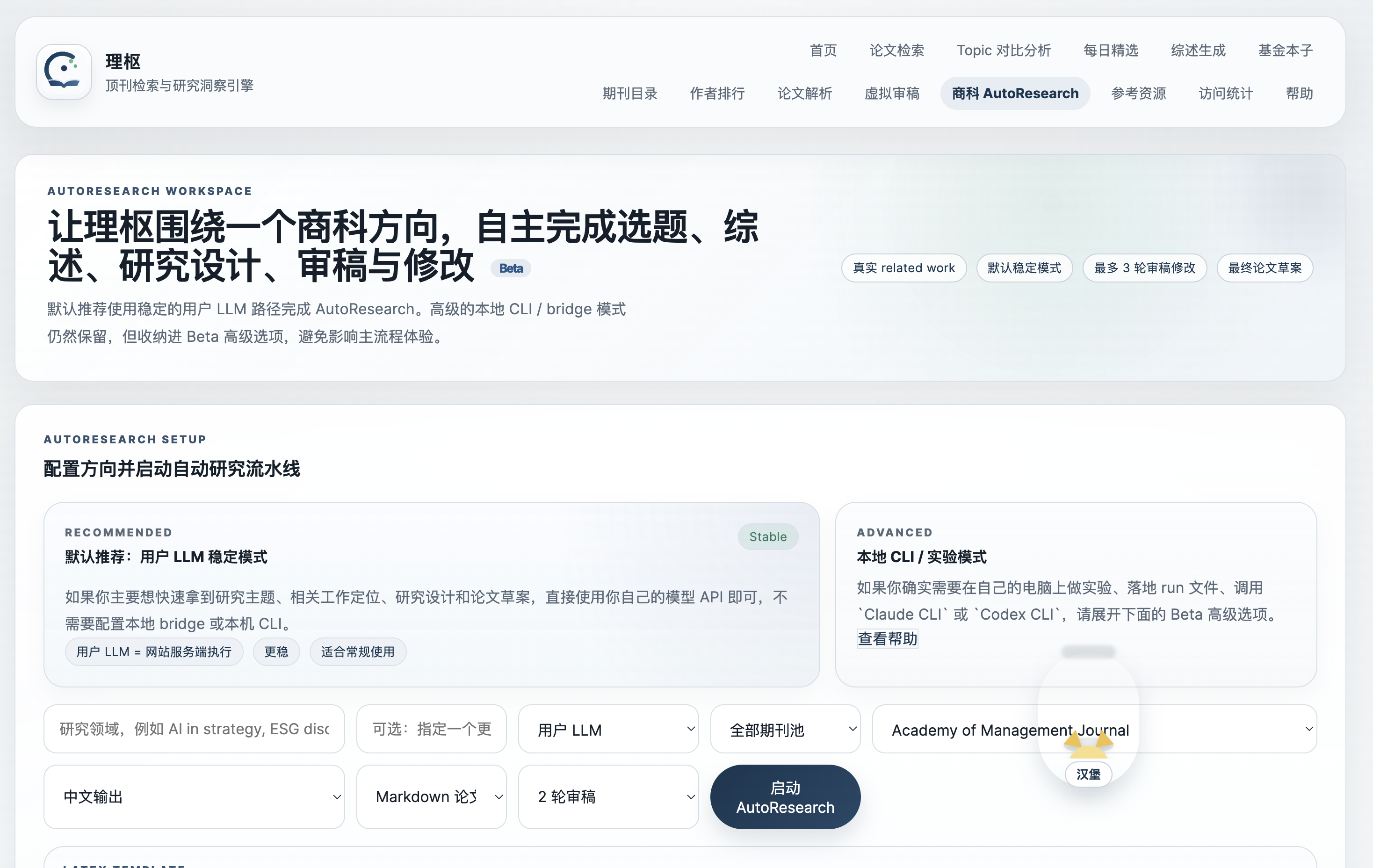}
\end{minipage}
\caption{Implemented AI-assisted workflow tools showing paper reading and AutoResearch orchestration}
\label{fig:workflow-tools-screenshot}
\end{figure}

The implemented interface now supports a broader set of researcher-facing tasks than the initial version of the artifact. In addition to journal-pool search and analytics, the system includes daily briefing views, topic-comparison dashboards, review-generation interfaces, paper-reading assistance, proposal drafting, and AutoResearch orchestration. Instead of emphasizing low-fidelity mockups, the report therefore highlights the implemented views that researchers actually interact with during search, synthesis, and writing support. Figures~\ref{fig:homepage-screenshot}--\ref{fig:workflow-tools-screenshot} serve not only as screenshots, but as evidence of how the artifact translates architectural intentions into concrete workflows.

\begin{figure}[t]
\centering
\begin{minipage}[t]{0.49\textwidth}
  \centering
  \includegraphics[width=\textwidth,height=0.34\textheight,keepaspectratio]{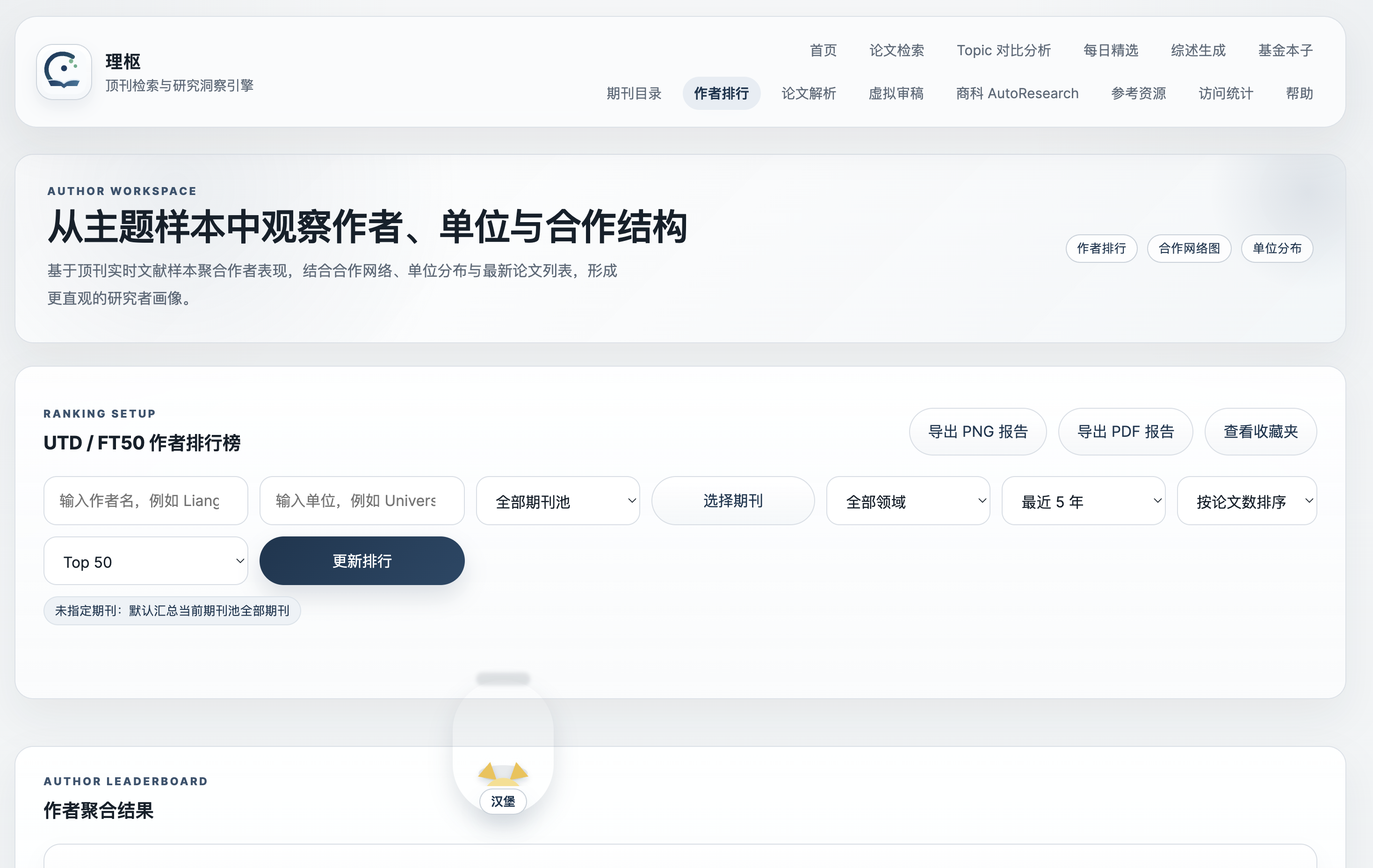}
\end{minipage}
\hfill
\begin{minipage}[t]{0.49\textwidth}
  \centering
  \includegraphics[width=\textwidth,height=0.34\textheight,keepaspectratio]{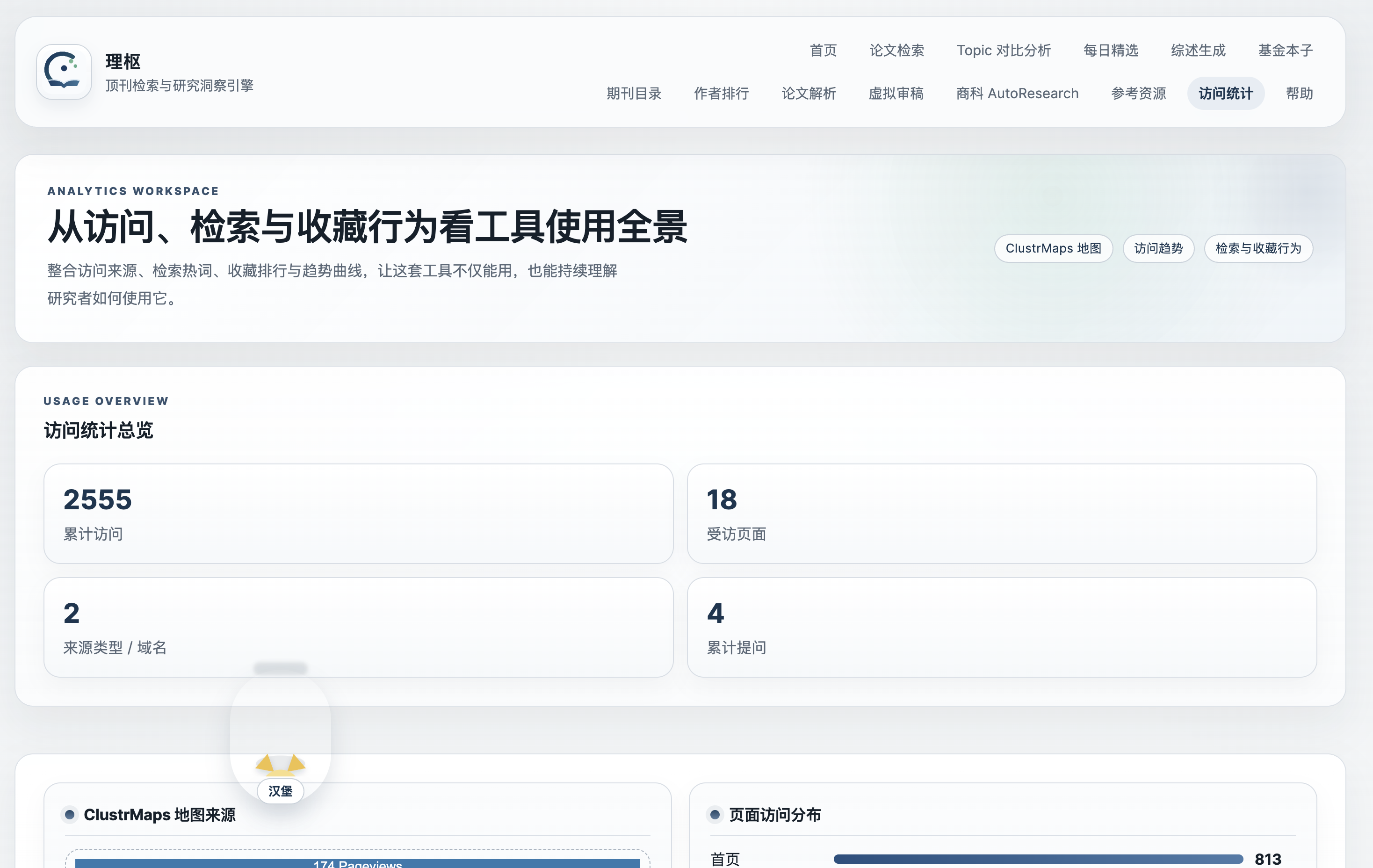}
\end{minipage}
\caption{Implemented author and analytics views showing author-network exploration and portal-level usage monitoring}
\label{fig:analytics-screenshot}
\end{figure}

Figure~\ref{fig:homepage-screenshot} illustrates the artifact's front-door logic: it foregrounds source transparency, journal-pool framing, and navigational shortcuts. Figure~\ref{fig:search-screenshot} shows the core retrieval surface, where live metadata retrieval, filtering, full-text preview, citation export, and hotspot analysis are tightly coupled in a single page. Figure~\ref{fig:comparison-daily-screenshot} extends the artifact into ongoing field monitoring through topic-comparison and daily-briefing views. Figure~\ref{fig:writing-screenshot} shows that the artifact now supports structured downstream writing tasks such as literature review generation and proposal drafting. Figure~\ref{fig:workflow-tools-screenshot} highlights AI-assisted reading and orchestration workflows, whereas Figure~\ref{fig:analytics-screenshot} captures both author-level exploration and portal-level usage monitoring. Together, these implemented views demonstrate that the artifact is not only technically functional but also structurally aligned with recurring scholarly work practices.

\section{Research Analytics Layer}
\label{sec:analytics}

The research analytics layer is the artifact’s most distinctive feature. It aims to support rapid scholarly orientation rather than formal bibliometric inference. In other words, the portal is designed less as a replacement for specialized bibliometric software and more as a decision-support layer that helps researchers recognize what is salient in a query result set quickly enough to guide the next step of reading, downloading, or theorizing.

\subsection{Hotspot Extraction}

Hotspot phrases are derived from titles and abstracts of the current result set. Candidate phrases are scored and then grouped into interpretable categories such as AI/Algorithms, Consumer/Market, Operations/Supply Chain, Finance/Accounting, and Strategy/Organization. When no native article keywords are available, the system falls back to rule-based extraction. When an LLM endpoint is configured, hotspot labels and quick takes can be rewritten into more readable summaries. The current interface also surfaces full-text availability signals such as open-access status, PDF availability, source priority, and in-page excerpt preview when an open version can be located.

\subsection{Distributional Analytics}

The portal computes a lightweight analytical layer from the current search result set:

\begin{itemize}[leftmargin=1.3em]
\item journal distribution,
\item category distribution,
\item year distribution,
\item keyword distribution,
\item top affiliations,
\item method signals,
\item top cited papers,
\item abstract and affiliation coverage.
\end{itemize}

These summaries provide immediate situational awareness for users exploring unfamiliar topics across multiple elite journals. This matters because management researchers often work across fragmented topical boundaries; a lightweight but interpretable analytic layer can materially reduce the cognitive effort required to move from article retrieval to exploratory synthesis.

\begin{figure}[t]
\centering
\begin{tikzpicture}
\begin{axis}[
  width=0.9\textwidth,
  height=5.2cm,
  ymin=0,
  xmin=1, xmax=10,
  xlabel={Recent 10-Day Window},
  ylabel={Cumulative Visits},
  axis line style={draw=accent},
  tick style={draw=accent},
  ymajorgrids=true,
  grid style={draw=linegray!60},
  xtick={1,2,3,4,5,6,7,8,9,10},
  xticklabels={D-9,D-8,D-7,D-6,D-5,D-4,D-3,D-2,D-1,D0},
  legend style={draw=none, fill=none, at={(0.02,0.95)}, anchor=north west}
]
\addplot[color=accent, thick, mark=*] coordinates {
  (1,4) (2,6) (3,8) (4,11) (5,13) (6,16) (7,20) (8,23) (9,27) (10,31)
};
\addlegendentry{Example cumulative traffic trend}
\end{axis}
\end{tikzpicture}
\caption{Illustrative cumulative analytics view used in the portal dashboard}
\label{fig:trend}
\end{figure}
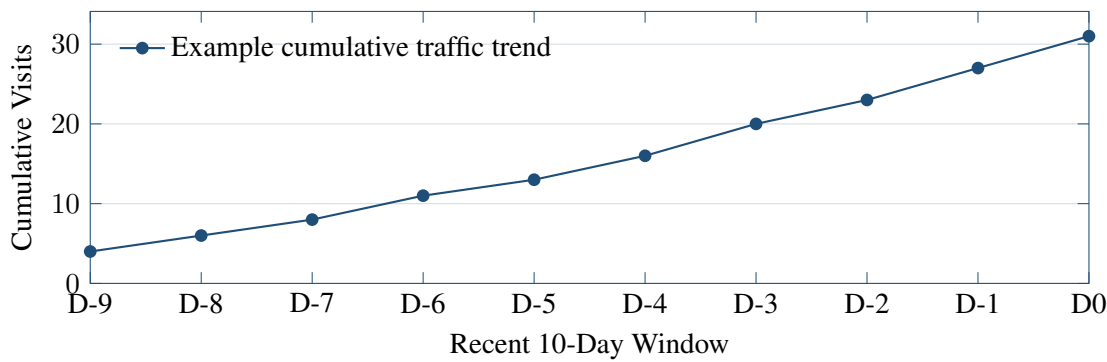

\subsection{Behavioral Analytics}

The artifact also records page visits, search keywords, and favorites events. These events are aggregated into top pages, source distributions, most frequent search terms, and most favorited papers. For public deployment, a ClustrMaps widget can be attached to the analytics page to visualize geographic traffic sources. Although these measures are operational rather than theoretical, they create a useful substrate for future evaluation studies by revealing how scholars actually interact with the artifact over time.

\begin{table}[t]
\centering
\caption{Analytics Objects and Intended Use}
\label{tab:analytics-objects}
\begin{tabularx}{\textwidth}{>{\raggedright\arraybackslash}p{0.23\textwidth} >{\raggedright\arraybackslash}p{0.29\textwidth} Y}
\toprule
\textbf{Analytics Object} & \textbf{Computation Basis} & \textbf{Primary Use} \\
\midrule
Hotspot phrases & Title and abstract phrase extraction with scoring & Rapid topic sensing within a query result set \\
Keyword distribution & Crossref subject fields or fallback extraction & Quick topical orientation \\
Top affiliations & Aggregated author affiliation strings & Institutional awareness and field mapping \\
Method signals & Heuristic text pattern detection & Rough understanding of empirical orientation \\
Top favorited papers & Aggregated user bookmark events & Monitoring researcher interest and repeated use \\
Visit trend and source mix & Pageview events and referrer classification & Artifact monitoring and public usage awareness \\
\bottomrule
\end{tabularx}
\end{table}

\section{Technical Architecture and Implementation}
\label{sec:implementation}

\subsection{Technology Stack}

The artifact uses Node.js for server-side routing and API orchestration, vanilla JavaScript for front-end logic, HTML/CSS for multi-page rendering, ECharts for result-set and analytics visualization, Supabase for free-tier persistence of usage analytics, and Render as a practical public deployment target. The live retrieval layer integrates Crossref, OpenAlex, Unpaywall, and optional CORE enrichment. The implementation deliberately avoids a heavier front-end framework in order to minimize setup friction and hosting complexity. This restraint is important for the artifact's design logic: a research support system is more likely to be adopted, replicated, and extended when the technical barrier to deployment remains low.

\subsection{Front-End and Interaction Design}

The front end is organized as a modular multi-page interface rather than a monolithic single-page application. This choice simplifies routing, page ownership, and public hosting while preserving a coherent user journey across home, journal browsing, live search, topic comparison, daily briefing, review generation, virtual review, paper reading, grant writing, AutoResearch orchestration, and analytics. Browser-side logic manages form state, URL synchronization, multi-journal filtering, favorites persistence, citation actions, and optional user-provided large-language-model configuration. From a software-engineering perspective, this structure improves maintainability by keeping task-specific rendering concerns loosely coupled while still allowing shared interface patterns across pages.

\subsection{Back-End Retrieval and Normalization}

The back end mediates between curated journal pools and live metadata retrieval. For each query, the service first resolves the relevant journal set, then issues bounded Crossref requests on a journal-by-journal basis, supplements records with OpenAlex where available, enriches full-text access candidates through Unpaywall and optional CORE lookup, normalizes returned records into a common schema, removes duplicates, applies filters, and computes analytical summaries. This orchestration layer is central to the artifact's contribution: it operationalizes prestige-based journal curation as a search boundary, thereby aligning retrieval behavior with the evaluative logic commonly used in business schools.

\subsection{Data Provenance and Source Integrity}

The artifact is intentionally conservative about data provenance. Journal pools are anchored in documented UTD-24 and FT50 sources, while paper-level content is derived from Crossref and OpenAlex records and exposed through DOI links and source labels. Open-access and full-text hints are derived from Unpaywall and optional CORE enrichment rather than opaque scraping. When article keywords are unavailable from native metadata, the system falls back to explicit heuristic extraction rather than presenting opaque or untraceable enrichment. Optional LLM support is treated as an interpretive and writing enhancement layer rather than a source of authoritative bibliographic truth.

\subsection{Persistence and Deployment}

One implementation challenge concerns analytics persistence on low-cost hosting. Some free web-service environments provide ephemeral file systems, which would reset local counters on redeploy. The artifact therefore supports two persistence modes:

\begin{enumerate}[leftmargin=1.3em]
\item local JSON storage for local development and quick tests,
\item Supabase-backed analytics persistence for stable public deployment.
\end{enumerate}

This decision is not only technical but also methodological. If the artifact is to be used as public research infrastructure, persistence reliability becomes part of the artifact’s practical validity. A deployable research artifact should not merely demonstrate conceptual fit; it should also remain stable enough for repeated public use, classroom demonstration, or lab-level extension.

\subsection{Citation and Export Support}

Each article object can be exported or cited through several output modes, including BibTeX, APA, MLA, Chicago, and plain text. CSV export and report export are also supported. Beyond search, the current artifact also exposes higher-level writing workflows such as review generation, virtual review simulation, proposal drafting, and AutoResearch-style planning. These functions position the artifact closer to an everyday research workbench than a demonstration-only prototype. In practice, they reduce a common break in scholarly workflow: the transition from discovering relevant papers to incorporating them into notes, bibliographies, analytical dashboards, and downstream writing tasks.

\section{Discussion}

The artifact contributes to design science and scholarly infrastructure in three ways. First, it demonstrates a domain-specific scholarly search artifact centered on elite business journals rather than undifferentiated bibliographic breadth. Second, it shows how real-source metadata, open-access enrichment, and lightweight analytics can be combined into a usable search-to-insight workflow without requiring proprietary infrastructure. Third, it extends that workflow into writing-support tasks such as review generation, proposal drafting, and AI-assisted reading while preserving traceable source boundaries. More broadly, the artifact suggests that there is substantial room for middle-layer scholarly systems: systems that do not compete with large bibliographic platforms on scale, but outperform them on domain alignment, transparency, and task focus.

At the same time, the artifact has important limitations. Coverage still depends on the quality and timeliness of Crossref, OpenAlex, and open-access availability services, which vary across journals and articles. The hotspot layer is optimized for interpretability rather than formal topic-model validity. Although the system can preview open-access text excerpts, it does not yet provide comprehensive full-text indexing or citation-network exploration. The higher-level writing modules are useful as research support, but they should be understood as scaffolding and synthesis aids rather than substitutes for scholarly judgment. Finally, the artifact has not yet undergone formal user evaluation with faculty or doctoral students. These limits should be interpreted as research opportunities rather than as disqualifying weaknesses: they point to the next design cycle needed to move from promising artifact to evaluated scholarly infrastructure.

\section{Implications and Future Work}

Several extensions are especially promising:

\begin{itemize}[leftmargin=1.3em]
\item \textbf{User evaluation:} task-based studies with doctoral students and faculty to assess search efficiency, interpretability, and perceived usefulness.
\item \textbf{Richer analytics:} topic modeling, co-author networks, institution collaboration views, and cross-pool comparison logic.
\item \textbf{Temporal intelligence:} rolling topic evolution across years, pools, and journals, enabling dynamic field monitoring.
\item \textbf{Submission support:} connecting observed topic concentration to outlet targeting guidance and journal-fit recommendations.
\item \textbf{Artifact generalization:} adapting the same architecture to adjacent domains such as finance, information systems, operations, or interdisciplinary journal pools.
\end{itemize}

Taken together, these directions suggest that the portal can evolve from a useful search utility into a broader research-infrastructure artifact: one that supports discovery, interpretation, and eventually strategic research planning.

\section{Conclusion}

Lishu is a focused research artifact for discovering, searching, and analyzing literature from elite business journals. By integrating UTD-24 and FT50 curation with live multi-source retrieval, open-access enrichment, analytical summaries, optional AI enhancement, citation utilities, and deployable analytics infrastructure, the artifact offers a practical response to the fragmented search workflows often faced by management researchers. The current system also demonstrates how a search artifact can evolve into a broader research workbench that supports reading, comparison, review generation, and proposal writing without abandoning source transparency. Although it remains an early-stage artifact rather than a finished scholarly platform, it already demonstrates a credible and extensible design pattern for business-school-oriented digital research infrastructure. For ICIS-style design and artifact discussions, the promise of the system lies not only in what it currently does, but in the evaluative and infrastructural research program it makes possible.

\section*{Acknowledgments}

The current public-facing artifact credits \textit{Designed by Chuang Zhao (ADM Lab)}. The implementation also integrates external data and infrastructure services, including Crossref, OpenAlex, Unpaywall, CORE, Supabase, Render, and ClustrMaps.

\end{document}